\documentclass[9pt,twocolumn,twoside]{osajnl}
\usepackage[ justification=justified]{caption} 
\usepackage{multirow}
\usepackage{titlesec}
\graphicspath{{images/}{../images/}}
\usepackage{tabularx}
\usepackage{physics}
\usepackage{bm}
\usepackage{bbm}
\usepackage{makecell} \usepackage[caption=false,subrefformat=parens,labelformat=parens]{subfig}
\usepackage{multibib}

\newtheorem{theorem}{Theorem}

\journal{optica} 
\setboolean{shortarticle}{false}
\usepackage{lineno}
\newcites{S}{References}
\title{Experimental composable key distribution 
 using discrete-modulated continuous variable quantum cryptography}

\author[1, *,$\dagger$]{Adnan A.E. Hajomer}
\author[2,3,$\dagger$]{Florian Kanitschar}
\author[1]{Nitin Jain}
\author[3]{Michael Hentschel}
\author[1]{Runjia Zhang}
\author[4]{Norbert Lütkenhaus}
\author[1]{Ulrik L. Andersen}
\author[3, 5]{Christoph Pacher}
\author[1,**]{Tobias Gehring}

\affil[1]{Center for Macroscopic Quantum States (bigQ), Department of Physics, Technical University of Denmark, 2800 Kongens Lyngby, Denmark}
\affil[2]{Vienna Center for Quantum Science and Technology (VCQ), Atominstitut, Technische Universität Wien, Stadionallee 2, 1020 Vienna, Austria}
\affil[3]{AIT Austrian Institute of Technology, Center for Digital Safety $\&$ Security, Giefinggasse 4, 1210 Vienna, Austria}
\affil[4]{Institute for Quantum Computing and Department of Physics and Astronomy, University of Waterloo, Waterloo, Ontario N2L 3G1, Canada}
\affil[5]{fragmentiX Storage Solutions GmbH, xista science park, Plöcking 1, 3400 Klosterneuburg, Austria}
\affil[$\dagger$]{These authors contributed equally}
\affil[*]{Corresponding authors: * aaeha@dtu.dk, ** tobias.gehring@fysik.dtu.dk}

\begin{abstract}
Establishing secure data communication necessitates secure key exchange over a public channel. Quantum key distribution (QKD), which leverages the principles of quantum physics, can achieve this with information-theoretic security. The discrete modulated (DM) continuous variable (CV) QKD protocol, in particular, is a suitable candidate for large-scale deployment of quantum-safe communication due to its simplicity and compatibility with standard high-speed telecommunication technology.
Here, we present the first experimental demonstration of a four-state DM CVQKD system, successfully generating composable finite-size keys, secure against collective attacks over a 20 km fiber channel with $2.3\times10^9$ coherent quantum states, achieving a positive composable key rate of $11.04\times10^{-3}$ bits/symbol. This accomplishment is enabled by using an advanced security proof, meticulously selecting its parameters, and the fast, stable operation of the system. Our results mark a significant step toward the large-scale deployment of practical, high-performance, cost-effective, and highly secure quantum key distribution networks using standard telecommunication components.

\end{abstract}

\setboolean{displaycopyright}{false}

\begin{document}

\maketitle
\section{Introduction}

Quantum key distribution (QKD)~\cite{BennettBrassard1984, Ekert1991} has emerged as a pivotal technology for secure communication, leveraging the principles of quantum mechanics to enable information-theoretic secure key exchange between two (or more) distant parties. Among the various approaches to QKD, continuous variable (CV) QKD is particularly notable for its compatibility with standard telecom technologies, allowing room temperature operation and high-rate secure key distribution~\cite{hajomer2024continuous, wang2022sub, pan2022experimental} over metropolitan distances compared to discrete-variable QKD ~\cite{li2023high, grunenfelder2023fast} which currently facilitates key exchange over higher channel attenuations than CV-QKD. This compatibility also facilitates miniaturization through photonic integration~\cite{hajomer2023continuous, zhangIntegratedSiliconPhotonic2019b}, and allows seamless integration with current telecom networks~\cite{eriksson2019wavelength, brunner2023demonstration, jain2024future}.

In the realm of CVQKD, Gaussian modulated protocols~\cite{Cerf2001, Grosshans2002, Grangier2002, Silberhorn2002} have traditionally dominated the field. These protocols use coherent states with Gaussian-distributed quadratures to encode key information, have fairly advanced security proofs~\cite{leverrier2015composable, pirandola2021composable}, and feature in all of the long-distance-record experiments for CVQKD (see Table~2 in Ref.~\cite{hajomer2024long}). Despite these advantages, Gaussian-modulated CVQKD protocols face significant implementation challenges. One major issue is the need for a large constellation of states to accurately approximate the continuous Gaussian distribution assumed in ideal security proofs. This requirement necessitates a high bit resolution for the digital-to-analog converter (DAC), which not only limits system speed but also complicates the integration of practical, coherent telecommunication components. This complexity makes it significantly harder to implement fast error-correction routines, thus becoming a major bottleneck preventing real-time execution of the complete protocol. Additionally, it also places a high demand on the rate from the quantum random number generator. Finally, even with a good approximation of continuous Gaussian modulation (through a large number of states in the constellation~\cite{jain2021practical}), a complete security analysis must account for the impact of discretization~\cite{Jouguet2012}. 

Discrete-modulated (DM) CVQKD~\cite{Ralph1999, Heid2006, Zhao2009} addresses these issues directly by using a finite set of quantum states, such as those from a quadrature phase shift keying (QPSK) alphabet. This approach simplifies the system implementation and makes it more accessible for real-world applications. Recent advancements in security analysis~\cite{Ghorai2019, Lin2019, Denys2021} have provided strong theoretical foundations for the composable finite-size security~\cite{Lupo2022, Kanitschar2023, Baeuml2023, PascualGarcia2024} of DM CVQKD. However, there has been a lack of experimental demonstrations validating the practical viability of distributing composable secure keys using DM CVQKD. 

In this article, we report the first experimental demonstration of DM CVQKD with composable finite-size security against collective attacks over a 20 km fiber channel. We achieved this using a CVQKD system implementing standard QPSK modulation and deploying an advanced composable finite-size security proof ~\cite{Kanitschar2023}. By carefully optimizing the system and ensuring high stability and high-speed operation, we achieved a positive composable key fraction of $11.04 \times 10^{-3}$ bits/symbol using a total of $N \approx 2.3 \times 10^{9}$ coherent quantum states with a security parameter of $\epsilon = 1\times 10^{-10}$. After implementing the full protocol stack, including classical error-correction and privacy amplification,  we obtained .94 Mbit of key material that, upon acceptance, is composably secure against independent, identically distributed (i.i.d.) collective attacks and ready for cryptographic tasks.

\begin{figure*}[t]
\centering
\includegraphics[width=0.8\linewidth]{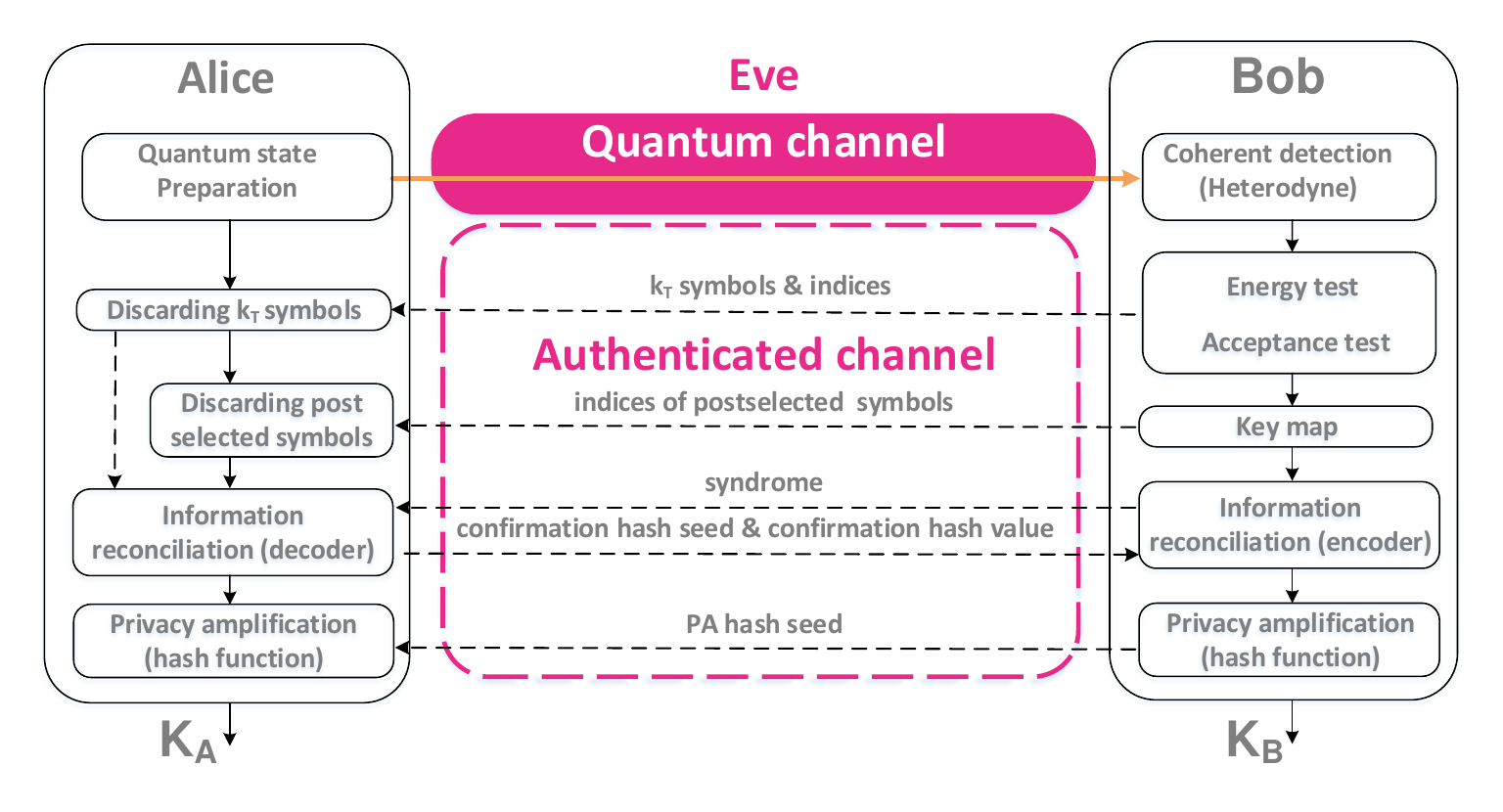}
\caption{\textbf{Discrete modulated continuous variable quantum key distribution protocol with composable security}. See the main text for the details.}
\label{fig:1}
\end{figure*}

\section{DM CVQKD Protocols and Composable Secure Key}\label{sec:Theory}
%Let us begin by introducing the executed DM CVQKD protocol followed by an overview of our security argument.

\subsection{Protocol Description}\label{subsec:ProtocolDescription}
Figure~\ref{fig:1} shows a flowchart of the prepare-and-measure DM CVQKD protocol with QPSK alphabet. The protocol steps read as follows:  
\begin{enumerate}
    \item[1] \textbf{\textit{State Preparation---}} Alice, using a random number generator, prepares one out of four coherent states $\ket{\alpha}$ with $\alpha \in \{|\alpha| e^{\frac{i\pi}{4}}, |\alpha| e^{\frac{3i\pi}{4}}, |\alpha| e^{\frac{5i\pi}{4}}, |\alpha| e^{\frac{7i\pi}{4}} \}$ according to a uniform distribution and sends it to Bob via the quantum channel controlled by Eve while keeping a record of the sent state in her classical register. We denote this classical two-bit register by $x_j$.

    \item[2] \textbf{\textit{State Measurement---}} Bob performs a heterodyne measurement on the received quantum state and determines the quadratures $q$ and $p$, which he stores as a complex number $y_j$ in his classical register.
\end{enumerate}
Steps $1$ and $2$ are repeated $N$ times.
\begin{enumerate}
    \item[3] \textbf{\textit{Energy Test---}} Once the quantum phase of the protocol is completed, Bob performs an Energy Test (see Supplementary information note \ref{APDX:secProofDetails} ) on $k_T<N$ randomly chosen symbols. The Energy Test (see Section \ref{subsec:SecurityArgument} and Theorem 2 in Ref. \cite{Kanitschar2023}) is an integral part of the security analysis that allows us to consider finite-dimensional Hilbert spaces, while still keeping a rigorous security statement. In case the Energy Test is passed, most of the weight of the received signals lies within a finite-dimensional Hilbert space, except with some small probability $\epsilon_{\text{ET}}$.  If the energy test is not passed, the protocol aborts.

    \item[4] \textbf{\textit{Acceptance Test---}} Alice discloses the data used for the Energy Test and Bob uses this information to determine statistical estimators for the observables used (see Supplementary Material note \ref{APDX:secProofDetails}). In case the Acceptance Test (see Theorem 3 in~\cite{Kanitschar2023}) is passed, the observed quantum states lie within the acceptance set, except with some small probability $\epsilon_{\text{AT}}$. If they do not lie within a predetermined acceptance set, the protocol aborts.

    \item[5] \textbf{\textit{Key Map---}} To determine a key string $\bar{z}$, Bob applies a key map on the remaining $n:= N-k_T$ symbols: he discretizes his measurement outcomes to elements in the set $\{0, 1, 2, 3, \perp\}$, where discarded symbols are mapped to $\perp$, allowing for postselection (see Refs. \cite{Lin2019, Kanitschar2021a} for details). The corresponding key map performed on each of the symbols reads
    \begin{equation}
            z_j(y_j) := \left\{
\begin{array}{ll}
0 \text{ if } 0 \leq \arg(y_j) <  \frac{\pi}{2} &\land~ \Delta_r \leq |y_j| \leq M, \\
1 \text{ if } \frac{\pi}{2} \leq \arg(y_j) <  \pi &\land~  \Delta_r \leq |y_j| \leq M, \\
2 \text{ if } \pi \leq \arg(y_j) <  \frac{3\pi}{2} &\land~  \Delta_r \leq |y_j| \leq M, \\
3 \text{ if } \frac{3\pi}{2} \leq \arg(y_j) <  2\pi &\land~  \Delta_r \leq |y_j| \leq M, \\
\perp &  \textrm{ otherwise,} \\
\end{array}
\right.
        \end{equation}
where $\Delta_r$ and $M$ are postselection parameters (see Section~\ref{subsec:SecurityArgument} for the meaning of $M$).

    \item[6] \textbf{\textit{Reverse Reconciliation---}} Bob uses the classical authenticated channel to send the syndrome of an error correcting code to Alice who corrects her ey string $\bar{x}$. 
    
    This is followed by Error Verification (sometimes called Confirmation): Alice and Bob compare hash-values of their key strings which they calculate using a randomly chosen hash function from a family of universal hash functions to confirm that all errors have been corrected successfully. Except with probability $\epsilon_{\text{EC}}$ they share an identical string afterwards.    

    \item[7] \textbf{\textit{Privacy Amplification---}} Finally, Alice and Bob turn their bit string into the secure key by applying again a randomly chosen hash function from a universal family. Except with some small probability $\epsilon_{\text{PA}}$ they finally hold the same secure key $K_A = K_B$.
\end{enumerate}

\subsection{Security Argument} \label{subsec:SecurityArgument}
In this work, we applied the composable security proof described in detail in Refs. \cite{Kanitschar2023, Kanitschar_Thesis_2022}. In the following, we give an overview of the idea and its application to the present experiment. 

Since ultimately, we want to use the generated secure key in larger cryptographic tasks, we aim for so-called composable security \cite{Renner2004, Renner2005}. Mathematically, this can be expressed by bounding the trace-distance between Alice's and Bob's shared state $\rho_{S_A S_B E}$ and the ideal state representing a uniformly distributed key shared between Alice and Bob that is completely decoupled from Eve, $\rho_{\text{ideal}} := \frac{1}{|\mathcal{K}|} \sum_{k \in \mathcal{K}} \ketbra{k} \otimes \ketbra{k}\otimes \rho_E$,
\begin{equation}
           \frac{1}{2} \left|\left| \rho_{K_A K_B E} - \rho_{\text{ideal}} \right|\right|_1 \leq \epsilon. 
\end{equation}
The security parameter $\epsilon > 0$ can be operationally interpreted as Eve's advantage in being able to distinguish the real from the ideal key. Practically $\epsilon$-security can be subdivided into $\epsilon_{\text{corr}}$-correctness and $\epsilon_{\text{secr}}$-secrecy. While correctness treats the case where the protocol does not abort but Alice and Bob hold a different key, secrecy covers the case where the protocol does not abort but Alice's and Bob's keys are not private. Note that the protocol is trivially secure if it aborts.

We will now discuss how security in the context of the present implementation is achieved. Alice prepares her quantum states and sends them to Bob via the insecure quantum channel, which is assumed to be controlled by Eve, as shown in Fig.~\ref{fig:1}. Thus, our security analysis can only rely on Bob's measurement results and authenticated information exchanged over the classical authenticated channel. In particular, we cannot assume a priori that the maximum photon number of the received quantum optical signals is bounded, but we perform a so-called Energy Test (Step 3 in our protocol). For the Energy Test, Bob discloses $k_T$ randomly chosen symbols via the classical channel and analyzes his measurements in those symbols according to Theorem 2 in Ref.~\cite{Kanitschar2023}: he picks a weight $w \in[0,1]$, a photon cutoff number $n_c$, a testing parameter $\beta_{\text{test}}>0$ and a number of allowed outliers $\ell_T$ and counts the number of symbols in which the measurement results $y_k$ lie outside a circle with radius $\beta_{\text{test}}$ in the phase space. In case this count exceeds a certain threshold given by the Energy Test Theorem, Bob aborts the protocol. The test is designed to fail except with some small probability $\epsilon_{\text{ET}}$ for states with a weight larger than $w$ outside the cutoff space.

Once the Energy Test is passed, Alice and Bob use the disclosed data to perform an Acceptance Test (see Theorem 4 in Ref.~\cite{Kanitschar2023}). Before the protocol execution, they have agreed on an expected channel behavior and have defined an acceptance set according to the Acceptance Testing theorem. This set contains all density matrices that are compatible with the expected channel behavior. Then, based on the disclosed measurement results, they determine statistical estimators for their considered observables and check if they lie within the acceptance set. While the test aborts on states outside this set, except with probability $\epsilon_{\text{AT}}$, the security analysis is conducted on the states within the acceptance set. Note that the separation into the Energy Test and the Acceptance Test is somewhat didactic: they can in fact be summarized into one single statistical test, giving rise to a set $\mathcal{S}^{\text{ET\&AT}}$.

However, the Acceptance Test requires bounded observables, which we cannot guarantee so far for our CVQKD protocol. We solve this problem by introducing an artificial detection range~\cite{Kanitschar2023}. We choose a number $M > 0$ that is smaller or equal to the actual detection range of the physical detector employed in our experiment and use this to define a finite region $\mathcal{M}:= \{ \gamma \in \mathbb{C}:~ |\gamma| < M \}$ in phase space that decides which measurements to keep ($y_k \in \mathcal{M}$) and which measurements to discard ($y_k \notin \mathcal{M}$). Note that for a tight implementation (see Supplementary Note \ref{APDX:DataAnalysis} for details), the finite detection range $M$ and the energy testing parameter $\beta_{\text{test}}$ are related. Thus, we chose $\beta_{\text{test}}$ such that the energy test passes most of the time for the expected channel behavior in the honest implementation. Additionally, we want to point out that the whole procedure can be simply viewed as postselection, as already facilitated by Step 5 of our protocol (see Section \ref{subsec:ProtocolDescription}).

Also, note that without further modifications, the described testing strategy will fail almost all the time, as only a single statistic is accepted. To address this problem, we slightly extend the set of accepted statistics, making the test more noise-robust. For each observable $X$, we quantify this extension by a parameter $t_X:= t_F \mu_X$, where $t_F$ is introduced for convenience. In Ref. \cite{Kanitschar2023} this was referred to as 'Non-unique acceptance testing`.

While, in addition to steps 1-5, the protocol also comprises error-correction and privacy amplification (Steps 6 and 7), those purely classical steps enter the security argument only in an abstract way in the form of security parameters $\epsilon_{\text{EC}}$, $\epsilon_{\text{PA}}$ and a leakage parameter $\delta_{\text{leak}}^{\text{EC}}$. The latter we can obtain from the practical setup. Thus, we do not have to discuss the specifics of those classical routines in detail here, but postpone their discussion to Section \ref{subsec:Postprocessing}. In this work, we aim for security against i.i.d. collective attacks. In this class of attacks, Eve is assumed to prepare a fresh ancilla state. Each of those ancillae may interact with each protocol round in an identical way and henceforth is stored in Eve's quantum memory until Alice and Bob have finished executing their protocol. This leads to the following security statement.

\begin{theorem}[\textbf{Security against i.i.d.\ collective attacks} \cite{Kanitschar2023}]\label{thm:SecurityStatement}
Let $\mathcal{H}_A$ and $\mathcal{H}_B$ be separable Hilbert spaces and let $\epsilon_{\mathrm{ET}}, \epsilon_{\mathrm{AT}}, \bar{\epsilon}, \epsilon_{\mathrm{EC}}, \epsilon_{\mathrm{PA}} > 0$. The objective QKD protocol is $\epsilon_{\mathrm{EC}} + \max\left\{\frac{1}{2}\epsilon_{\mathrm{PA}}+\bar{\epsilon}, \epsilon_{\mathrm{ET}}+\epsilon_{\mathrm{AT}} \right\}$-secure against i.i.d.\ collective attacks, given that, in case the protocol does not abort, the secure key length $\ell$ is chosen to satisfy
\begin{equation}\label{eq:KRexpression}
\begin{aligned}
    \frac{\ell}{N} \leq \frac{n}{N} &\left[ \min_{\rho \in \mathcal{S}^{\mathrm{E\&A}}} H(X|E')_{\rho} - \Delta(w) - \delta(\bar{\epsilon}) \right]     \\
    & - \delta_{\mathrm{leak}}^{\mathrm{EC}} - \frac{2}{N} \log_2\left( \frac{1}{\epsilon_{\mathrm{PA}}} \right),
\end{aligned}
\end{equation}
where $\delta^{\mathrm{EC}}_{\mathrm{leak}}$ takes the classical error correction cost into account, $\Delta(w) := \sqrt{w} \log_2(|Z|) + (1+\sqrt{w}) h\left( \frac{\sqrt{w}}{1+\sqrt{w}} \right)$, $\delta(\bar{\epsilon}) := 2 \log_2\left( \mathrm{rank}(\rho_X)+3 \right) \sqrt{\frac{\log_2\left(2/\bar{\epsilon} \right)}{n}}$, $\mathcal{S}^{\mathrm{E\&A}}$ contains all states that pass both the Energy Test and the Acceptance Test except with probability $\epsilon_{\mathrm{ET}}+\epsilon_{\mathrm{AT}}$ and $n:= N - k_T$. 
\end{theorem}
By $H$, we mean the conditional von Neumann entropy; by $|Z|$, we denote the number of different key map elements that are not discarded during post-selection, which for the present protocol is $4$ and $\bar{\epsilon}$ is a smoothing parameter that appears in the security argument (see Ref. \cite{Kanitschar2023} for details). 
Note that the security statement in Theorem \ref{thm:SecurityStatement} can be split into correctness and secrecy, with corresponding parameters
$\epsilon_{\mathrm{corr}} = \epsilon_{\mathrm{EC}}$ and $\epsilon_{\mathrm{secr}} = \max\left\{\frac{1}{2}\epsilon_{\mathrm{PA}}+\bar{\epsilon}, \epsilon_{\mathrm{ET}}+\epsilon_{\mathrm{AT}} \right\}$. 

Before we address the optimization problem in Eq. (\ref{eq:KRexpression}), let us briefly comment on $\Delta(w)$. Since numerical security proofs cannot handle infinite-dimensional systems, for the evaluation of the key rate formula given in Eq. (\ref{eq:KRexpression}), we have to work with a numerical cutoff. However, to avoid simply assuming such a cutoff, we employ the dimension-reduction method~\cite{Upadhyaya2021} (using the improved correction term from Ref.~\cite{UpadhyayaThesis2021}) which relates the infinite-dimensional evaluation to a finite-dimensional cutoff representation at the cost of introducing said correction term $\Delta(w)$ which is a function of the weight $w$ we chose for the Energy Test. We will discuss the choice of $w$ and its interplay with other parameters in Section \ref{sec:Results}. 

The remaining task in the security analysis is to solve the optimization problem in Eq. (\ref{eq:KRexpression}), which can be cast into a semi-definite program (SDP) with a non-linear objective function \cite{Lin2019, Upadhyaya2021, Kanitschar2023}. We approach this minimization problem with a two-step process \cite{Coles2016, Winick2018}. First, we iteratively solve a linearized version of the problem using the Frank-Wolfe algorithm \cite{FrankWolfe1956}, obtaining an upper bound on the secure key rate. Second, we convert this upper bound into a lower bound, using the SDP duality theory. Finally, we relax the problem by taking numerical imprecisions into account, and obtain a reliable lower bound on the secure key rate. For details about the optimization problem, we refer the reader to Supplementary Note  \ref{APDX:secProofDetails}.

 \begin{figure*}[t]
\centering
\includegraphics[width=\linewidth]{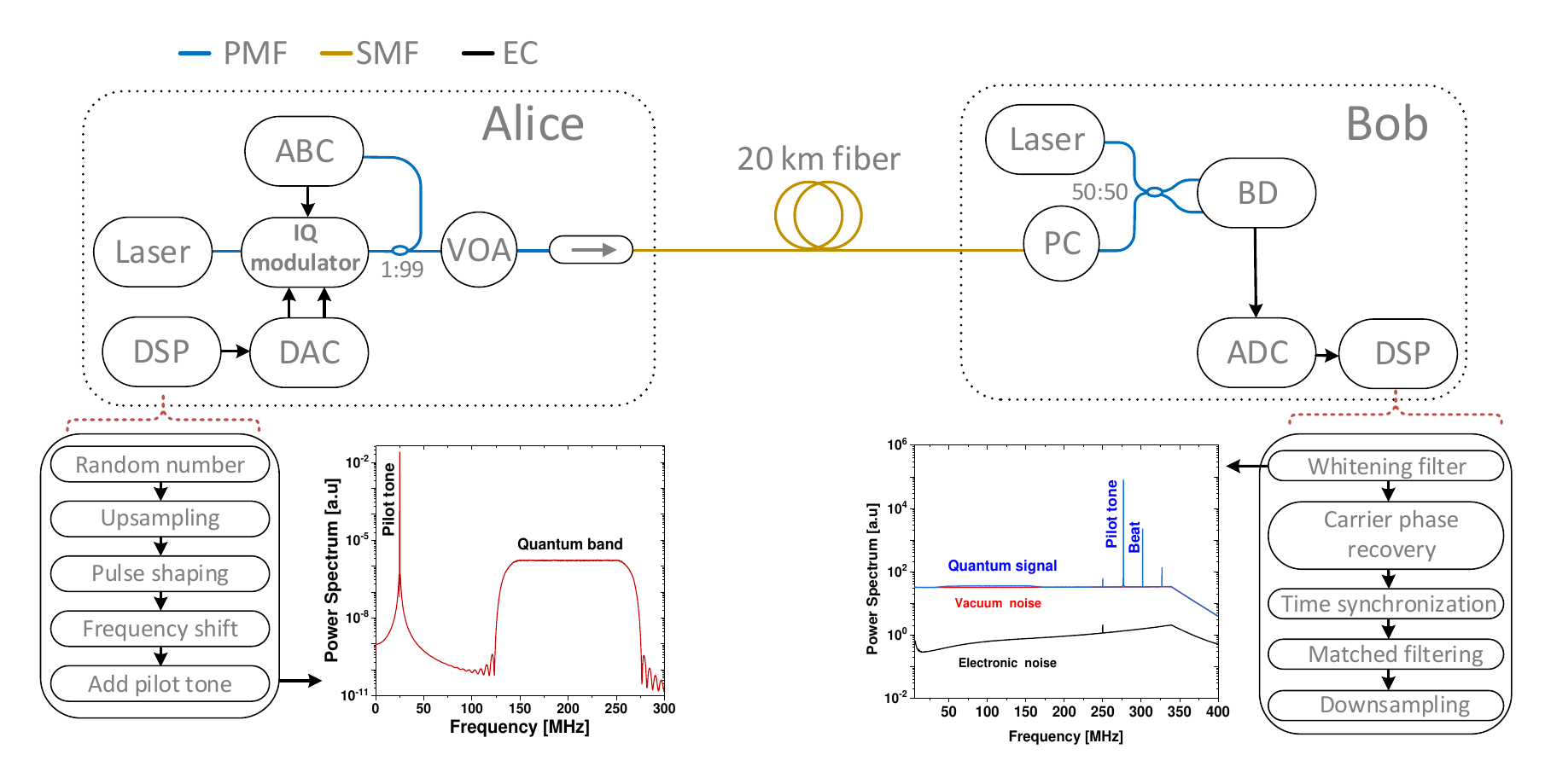}
\caption{\textbf{DM CVQKD set-up}. Schematic of the quantum key distribution (QKD) system, detailing all key components and digital signal processing (DSP) modules. At Alice's side: CW laser (continuous wave laser), IQ modulator (in-phase and quadrature modulator), VOA (variable optical attenuator), Faraday isolator (indicated by the arrow), DAC (digital-to-analog converter) and ABC (automatic bias controller). At Bob's side: CW laser used as a local oscillator (LO), BD (balanced detector),  polarization controller (PC), and ADC (analog-to-digital converter). Alice's and Bob's stations utilize polarization-maintaining fiber (PMF) components, while the quantum channel is a spool of standard single-mode fiber (SMF). All electronic connections (EC) are represented by black lines. } 
\label{fig:2}
\end{figure*}

\section{CVQKD System implementation}\label{sec:Implementation}

In Fig.~\ref{fig:2} we show a schematic of our CVQKD system, which uses advanced digital signal processing (DSP) and simple optical modules to facilitate secure key exchange between a sender (Alice) and a receiver (Bob) with composable security against collective attacks. We refer to this system as a digital CVQKD system~\cite{hajomer2024long,jain2021practical}, as it simplifies the optical subsystem by integrating hardware functions into the DSP module. The following subsections provide thorough details on both the optical subsystem and the DSP modules.
 
\subsection{Sender} 
Alice began the signal generation process by creating a digital waveform using the DSP module, as shown in the bottom left corner of Fig.~\ref{fig:2}. The four coherent state amplitudes, denoted by $\alpha_i = q_i + \iota p_i$, were formed by drawing real ($q_i$) and imaginary ($p_i$) parts from a binary sequence and scaling them: $q_i, p_i \in \frac{1}{\sqrt{2}}\times\text{Uniform}\{-1,1\}$. The sequence was produced by a pseudo-random number generator, which was used to simplify the implementation. However, it can be swapped out against a quantum random number generator (and it should) for the security statement to hold. The symbols were drawn at a rate of 125 MBaud and then upsampled to match the DAC sampling rate of 1 GSample/s. Digital pulse shaping was applied using a root-raised cosine (RRC) filter with a roll-off factor of 0.2, creating a band-limited baseband signal $m(t)$. To achieve single sideband modulation, the baseband signal $m(t)$ was frequency shifted by $f_c = 200$ MHz, resulting in a single sideband passband signal $\bar{m}(t)= m(t)e^{2i\pi f_c t}$. A 25 MHz pilot tone was frequency multiplexed to this upconverted signal for carrier phase recovery. The spectrum of the generated digital waveform uploaded to the DAC is shown in the bottom left corner of Fig.~\ref{fig:2}. 

Alice's optical subsystem featured a 1550 nm continuous wave (CW) laser with a linewidth of 100 Hz. A commercial off-the-shelf in-phase and quadrature (IQ) modulator, driven by the DAC, displaced the coherent states in the phase space. The IQ modulator operated in an optical single sideband carrier suppression mode, with bias voltages controlled by an automatic bias controller (ABC). To ensure non-orthogonal coherent states at the quantum channel input, the amplitude of the generated states was attenuated using a variable optical attenuator (VOA). Additionally, to prevent Trojan horse attacks, an isolator, in the figure marked by an ellipse with an arrow, was placed before the quantum channel made of a 20 km standard single-mode fiber (SMF). 

\subsection{Receiver} 
To decode the quantum information, Bob used a digital coherent receiver, consisting of an optical subsystem and a DSP module, shown on the right side of Fig.~\ref{fig:2}. In the optical subsystem, radio frequency heterodyne detection was performed by overlapping the received quantum signal with a local oscillator (LO) at a balanced beamsplitter. The LO was generated from an independent free-running CW laser with a frequency offset of approximately 302 MHz relative to Alice's laser. Because this frequency offset exceeded half the bandwidth of the quantum signal, the amplitude and phase quadratures were measured concurrently using a single balanced detector (BD) with a bandwidth of approximately 350 MHz. A manual polarization controller (PC) was used to align the polarization of the electro-magnetic field of the light at the output of the quantum channel to the fixed polarization of the LO by maximizing the interference visibility. The detected signal was then digitized using a 1 GSample/s analog-to-digital converter (ADC) and synchronized to the DAC with a 10 MHz reference clock.

Bob's DSP module began with a whitening filter to remove correlations in the received symbols caused by the non-flat response of the BD. The filter coefficients were the inverse frequency response of the BD, computed from vacuum noise.  Figure~\ref{fig:2} (the bottom left corner) shows the spectrum of the received signal after the whitening filter. The next step was carrier phase recovery, which included frequency estimation using the pilot tone and phase estimation using an unscented Kalman filter (UKF)~\cite{chin2021machine}. The propagation delay of the fiber channel and various electronic components was estimated by cross-correlating reference and receiver samples. Finally, the quantum symbols were recovered through matched filtering and downsampling to the symbol rate of 125 MBaud. 

\subsection{System calibration and measurements}
In DM CVQKD, optimizing the average amplitude of the generated coherent states ensemble is crucial for maximizing the secure key rate for a given channel loss. This process began with theoretically determining the optimal average amplitude for a quantum channel with a physical loss of approximately 5 dB and was found to lie between 0.68 and 0.72  (see section~\ref{sec:Results} for more details).

We then conducted back-to-back measurements, connecting the sender and receiver directly with a short fiber patch cord, to calibrate the system. Using the VOA and adjusting the DAC driving voltage, we fine-tuned the average amplitude of the coherent state ensemble to 0.71.

After calibrating the average amplitude of the generated states ensemble, we connected the quantum channel and performed three consecutive measurements: quantum signal measurement, vacuum noise measurement (LO laser on, Alice's laser off), and electronic noise measurement (LO laser off, Alice's laser off). These measurements were conducted automatically using a Python-based framework, eliminating the need for user intervention. To expedite offline DSP using parallel processing, each measurement was divided into frames of $10^7$ ADC samples, with a total of $2 \times 10^{10}$ samples collected for each type of measurement. Let us remark that our system operated in a non-paranoid scenario, assuming some loss and noise were beyond Eve's control. Therefore, the average amplitude of 0.71 was calculated considering a trusted receiver efficiency of $68\%$. Following these optical measurements and offline DSP, the remainder of the DM CVQKD protocol was performed.

\section{Post-Processing} \label{subsec:Postprocessing}

The post-processing phase is responsible for transforming the DSP-processed measurement data into a secure key. As illustrated in Fig.~\ref{fig:1}, this phase encompasses several tasks, already defined and described in Section~\ref{sec:Theory}. These tasks require communication over a classical channel, which is assumed to be error-free and authenticated to prevent man-in-the-middle attacks. Information-theoretic secure authentication is achieved by applying a universal hash function to all messages exchanged between Alice and Bob and using a small pre-shared key, which is periodically replenished with secure quantum keys.

\subsection{Energy Test and Acceptance Test}

The first step, Energy and Acceptance Tests, involves disclosing $k_T$ symbols. We performed the security analysis for three different values of $k_T$, i.e., $0.4\times N$, $0.45\times N$ and $0.5\times N$  (recall $N\approx 2.3\times10^9$). After these tests, the disclosed symbols were used to estimate parameters, such as the signal-to-noise ratio (SNR), which are critical for subsequent tasks.  Afterwards, the disclosed symbols were discarded.

\subsection{Key Mapping \& Post Selection}

To implement key mapping on Bob's measurements, radial post selection was employed with the parameter $M$ set to 3.889 natural units (NU) (see Supplementary Note~\ref{APDX:DataAnalysis} for the definition of NU), while $\Delta_r$ was chosen within the range of 0.3 to 0.7 NU. These parameters are vital for the error-correction step, as they allow control over the SNR by discarding data below $\Delta_r$ and above $M$. The key mapping step concludes with the disclosure of a fraction $r_{\perp}$ of the discarded symbols. 

\subsection{Reverse Reconciliation}
Following key mapping, Alice and Bob perform reverse reconciliation using low-density parity-check (LDPC) codes, which operate close to the Shannon limit at low SNRs. 

We have created a collection of LDPC codes with a constant block size of 512'000 bits for the binary symmetric channel (BSC) and code-rates adapted to the relevant range of SNRs (see Table~\ref{tab:PS}). 

We characterize each LDPC code by its code rate $R=\frac{L_\mathrm{LDPC}-L_\mathrm{syn}}{L_\mathrm{LDPC}}$, with the block length $L_\mathrm{LDPC}$ and the syndrome length $L_\mathrm{syn}$, and its SNR-threshold which we define to be the SNR where the frame error rate $FER=0.5$ using a maximum of 200 decoder iterations. The information leakage $\mathrm{EC}_\mathrm{leak}$ due to error-correction is calculated based on the number of corrected blocks $B_\mathrm{cor}$ and the number of failed blocks $B_\mathrm{fail}$: 

\begin{equation}\label{eq:ECLeak}
\begin{aligned}
\mathrm{EC}_\mathrm{leak} = &\frac{B_\mathrm{cor}}{B_\mathrm{cor}+B_\mathrm{fail}} \times \left(1-R \right) + \\
& \frac{mathrm{fail}}{B_\mathrm{cor}+B_\mathrm{fail}} \times \left(\mathrm{QRE} - \Delta(w) - \delta(\bar{\epsilon})\right),
\end{aligned}
\end{equation}
where $\mathrm{QRE}:= \min_{\rho \in \mathcal{S}^{\mathrm{E\&A}}} H(X|E')_{\rho}$ denotes the entropy term in Eq. (\ref{eq:KRexpression}),  $\Delta(w)$ is a weight correction term, and $\delta(\bar{\epsilon})$ is the correction due to applying the asymptotic equipartition property of the complete input block after post-selection. The first term represents the total information content in the syndromes of all corrected blocks, while the second term accounts for the complete information of all failed blocks. Even when a failed block is fully disclosed, the information leakage cannot exceed the block's contained information.

Alice and Bob divide their string of key mapped data symbols into blocks of 512'000 bits which matches the fixed block size $L_\mathrm{LDPC}$ of the LDPC codes. The LDPC code with the highest threshold below the estimated SNR value is selected for correction. Using the parity check matrix of the selected LDPC code Bob calculates and sends the syndrome of each block to Alice, who corrects her data block. We set the maximum number of LDPC decoder iterations to 200, which gives a good compromise between FER achieved and run-time. Our throughput optimized decoder converges typically in 20-100 decoder iterations, taking 1 ms per iteration on a single core of a 2.8 GHz AMD EPYC 7402P CPU. This corresponds to a throughput of roughly $5\times 10^6$ to $25\times 10^6$ corrected bits/s.

Table~\ref{tab:PS} illustrates the impact of the postselection parameter $\Delta_r$ on the SNR, the error-correction performance, and the leakage. For example, using the same LDPC code (see e.g. code $R=5\%$) with increasing $\Delta_r$ will result in a lower  thus reducing the leakage $\mathrm{EC}_\mathrm{leak}$. But at the same time, due to the rising SNR, the efficiency for corrected blocks drops, so the average efficiency (Eq. \ref{eq:beta}) still decreases in spite of lower FER.

\begin{equation}\label{eq:beta}
\bar{\beta}=\frac{1}{B_\mathrm{tot}} \sum_{k=1}^{B_\mathrm{tot}} \beta_\mathrm{k}=\frac{1}{B_\mathrm{tot}} \sum_{k=1}^{B_\mathrm{tot}} \frac{R}{I_\mathrm{AB,k}},
\end{equation}
where $B_\mathrm{tot}$ is the total number of blocks, $I_\mathrm{AB,k}$ is the mutual information for the $k^{th}$ block and using a code rate of zero for failed (and disclosed) blocks.
Moreover, due to the higher number of discarded symbols, the total secure key fraction drops (see Figure \ref{fig:Results_rPS}). This highlights the critical trade-off between these parameters to maximize the final secure key length.

\subsection{Error Verification (Confirmation) with Polynomial Hashing}
After error correction, Alice randomly selects a hash function from a family of polynomial universal hash functions that map from 512'000 to 96 bits and sends the function index to Bob. Then Alice and Bob use this function to calculate a hash value for each data block. Alice sends her hash value to Bob which compares Alice's and his hash value. If the hash values are different, reconciliation has failed and Bob discloses his data block. 

\subsection{Privacy Amplification}
Finally, all blocks are concatenated to form one large block. Alice computes the length of the final secure key based on Eq.~\ref{eq:KRexpression}, depending on the total security parameter $\epsilon$ and the information leakage from error-correction. 

Alice randomly selects a hash function from the family of universal Toeplitz hash functions~\cite{fung2010practical}, that map from the length of the input block to the length of the final secure key, and sends (the first row and column of) the Toeplitz matrix to Bob. 

Alice and Bob each apply this hash function and obtain the secure key.

\begin{table}[t]
\centering
\caption{\textbf{Effect of post-selection on error-correction with test ratio 40\%}. Post-selection parameter: $\Delta_r$, fraction of discarded symbols due to post-selection: $r_{\perp}$, signal-to-noise ratio: SNR, code rate: $R$,  frame error rate: FER, leakage: $\mathrm{EC}_\mathrm{leak}$.}
\resizebox{1\hsize}{!}{
\begin{tabular}{|c|c|c|c|c|c|}
\hline
\multicolumn{1}{|c|}{$\Delta_r$, NU} & \multicolumn{1}{c|}{$r_{\perp}$, \%} & \multicolumn{1}{c|}{SNR} & \multicolumn{1}{c|}{$R$, \%} & \multicolumn{1}{c|}{FER, \%}  & $\mathrm{EC}_\mathrm{leak}$, bits/symbol     \\ \hline
\multicolumn{1}{|c|}{0}    & \multicolumn{1}{c|}{0}    & \multicolumn{1}{c|}{0.0944} & \multicolumn{1}{c|}{3.5} & 0.018  & 1.9298 \\
\multicolumn{1}{|c|}{0.30} & \multicolumn{1}{c|}{7.70} & \multicolumn{1}{c|}{0.1027} & \multicolumn{1}{c|}{4} & 0.020 & 1.7721 \\ 
\multicolumn{1}{|c|}{0.35} & \multicolumn{1}{c|}{10.33} & \multicolumn{1}{c|}{0.1057} & \multicolumn{1}{c|}{4.5} & 0.445 & 1.7124 \\ 
\multicolumn{1}{|c|}{0.40} & \multicolumn{1}{c|}{13.28} & \multicolumn{1}{c|}{0.1091} & \multicolumn{1}{c|}{4.5} & 0.021 & 1.6562 \\ 
\multicolumn{1}{|c|}{0.45} & \multicolumn{1}{c|}{16.50} & \multicolumn{1}{c|}{0.1130} & \multicolumn{1}{c|}{5} & 3.997 & 1.5866 \\ 
\multicolumn{1}{|c|}{0.50} & \multicolumn{1}{c|}{19.96} & \multicolumn{1}{c|}{0.1174} & \multicolumn{1}{c|}{5} & 0.045 & 1.5207 \\ 
\multicolumn{1}{|c|}{0.55} & \multicolumn{1}{c|}{23.62} & \multicolumn{1}{c|}{0.1222} & \multicolumn{1}{c|}{5} & 0.024 & 1.4511 \\ 
\multicolumn{1}{|c|}{0.60} & \multicolumn{1}{c|}{27.43} & \multicolumn{1}{c|}{0.1275} & \multicolumn{1}{c|}{5.6} & 0.075 & 1.3700 \\ 
\multicolumn{1}{|c|}{0.65} & \multicolumn{1}{c|}{31.37} & \multicolumn{1}{c|}{0.1332} & \multicolumn{1}{c|}{6} & 3.462 & 1.2907 \\ 
\multicolumn{1}{|c|}{0.70} & \multicolumn{1}{c|}{35.37} & \multicolumn{1}{c|}{0.1394} & \multicolumn{1}{c|}{6} & 0.028 & 1.2149 \\ \hline
\end{tabular}
\label{tab:PS}
}
\end{table}

\begin{figure*}[t]
\centering
\includegraphics[width=\linewidth]{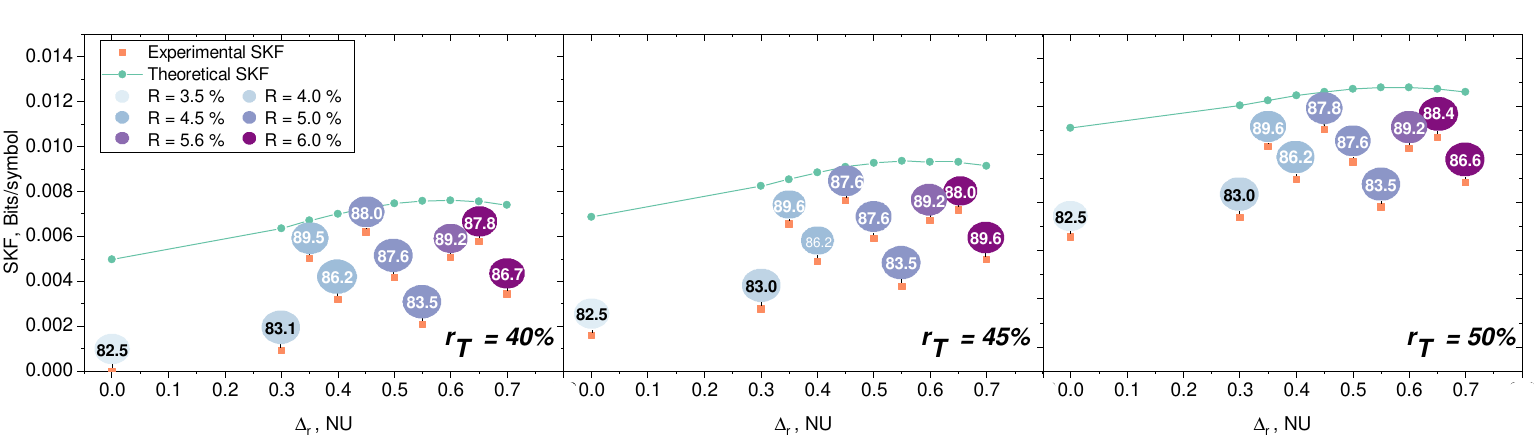}
\caption{ \textbf{Secure Key Fraction (SKF $=\ell / N$) vs. radial postselection parameter for different testing ratios $r_T$}. The green curves represent the theoretical SKF for assumed (constant) $92\%$ error-correction efficiency, the orange dots illustrate the secure key fraction for the experimental data. The numbers in the balloons state the average efficiency $\bar{\beta}$ of LDPC codes used for error-correction, while the color of the balloons indicates the code rate $R$.\label{fig:Results_rPS}}
\end{figure*}

\section{Results}\label{sec:Results}
Before presenting the results, it is essential to discuss and specify the parameters used in this work and to comment on the application of the security argument. 

Recall that the security argument behind Theorem \ref{thm:SecurityStatement} is based on comparing the observations with the honest implementation of the protocol, followed by a hard accept or abort decision. In our proof-of-concept experiment, we characterized the expected behavior in the non-adversarial scenario experimentally and chose appropriate quantities for both the Acceptance and the Energy Test. In particular, we chose $t_X = \mu_X$ (so, $t_F=1$) in the Acceptance Testing theorem (see Ref. \cite{Kanitschar2023} or Theorem \ref{Thm:ParameterEstimation}). For a full protocol run, the observed quantities must be processed as described by the tests. In case the observations pass the Energy Test and lie within the acceptance set, the terms $\min_{\rho \in \mathcal{S}^{\mathrm{E\&A}}} H(X|E')_{\rho}- \Delta(w)$ in the security statement are equal to the pre-calculated value, otherwise the protocol aborts. Thus, the computationally expensive key rate calculation is performed beforehand, during the characterization of the system, and is no bottleneck for the protocol execution. The generalization of recently published variable-length security arguments for discrete-variable QKD  \cite{Tupkary2024} to the continuous variable regime could ease application and improve key rates further.

In this work, we aim for a total composable security parameter of $\epsilon = 1 \times 10^{-10}$. The connection between the security parameters of the sub-protocols and the total security parameter is detailed in Theorem~\ref{thm:SecurityStatement}. Additionally, we need to take the security parameter of the Random Number Generator into account. However, following the idea of composable security, this simply adds to the existing security statement from Theorem~\ref{thm:SecurityStatement}. Consequently, we obtain $\epsilon = \epsilon_{\text{RNG}} + \epsilon_{\mathrm{EC}} + \max\left\{\frac{1}{2}\epsilon_{\mathrm{PA}}+\bar{\epsilon}, \epsilon_{\mathrm{ET}}+\epsilon_{\mathrm{AT}} \right\}$. Our approach was to approximately balance both terms in the maximum expression for the total security parameter, setting, $\frac{1}{2}\epsilon_{\text{PA}} + \bar{\epsilon} = \epsilon_{\text{ET}} + \epsilon_{\text{AT}}$.

The security parameter of the error-correction routine ($\epsilon_{\text{EC}}$) is related to error-verification and depends on the length of the corrected bitstring and the length of the hash tag ($b_\text{EV}$) created (see Theorem 2 in \cite{Johansson94polyhashing}):
\begin{align}
    \epsilon_{\text{EC}} = 2^{-b_{\mathrm{EV}}} \times \left\lceil \frac{L_\mathrm{LDPC}}{b_{\text{EV}}} \right\rceil \times \left\lceil \frac{n(1-r_{\perp})}{L_\mathrm{LDPC}}\right\rceil.
\end{align}
Here, the leaked information is bounded by the length of the hash tag $b_{\text{EV}}$, which we set to  $96$ bits. The security parameter for the privacy amplification routine ($ \epsilon_{\text{PA}}$) is given by the leftover hashing lemma \cite[Lemma 5.6.1]{Renner2005} used in the security proof (see Section \ref{subsec:SecurityArgument}), which ensures that the security parameter decreases exponentially as the key is shortened relative to $H_{\text{min}}^{\epsilon'}(X|E')_{\rho}$. Denoting the difference between the final key and the key length given by this entropic quantity by $b_{\text{PA}}$, we obtain the relation
\begin{equation}
    \epsilon_{\text{PA}} \leq 2^{-\frac{b_{\text{PA}}}{2}}.
\end{equation}
Due to this exponential relation, those security parameters can be made almost arbitrarily small. The choice of the security parameters linked to the statistical tests, $\epsilon_{\text{ET}}$ and $\epsilon_{\text{AT}}$ is discussed in detail in Supplementary Note \ref{APDX:secProofDetails}. Notably, the choice of $\epsilon_{\text{AT}}$ can significantly impact the secure key rates. Thus, we aim to choose this parameter as large as possible. Finally the smoothing parameter $\bar{\epsilon}$ is a `virtual' parameter that appears in the security proof and comes with a correction term. As $\epsilon_{\text{PA}}$ can be reduced at low cost, $\bar{\epsilon}$ was also chosen to be as large as possible. The selected security parameters are summarized in Table \ref{tab:secParams}. These represent upper bounds, ensuring a total security parameter of $10^{-10}$, though actual values may vary depending on other quantities (e.g., $\epsilon_{\text{EC}}$ on the key length $\ell$) and therefore can be smaller.

To determine the optimal coherent state amplitude ($|\alpha|$), we first measured the system detector parameters $\eta_D$ and $\nu_{\text{el}}$, the physical loss $\eta$ of the channel and the excess noise $\xi$ (see Supplementary Note  \ref{APDX:DataAnalysis}) in a non-adversarial scenario. Then, we numerically simulated the achievable secure key rates assuming a Gaussian channel and estimated the optimal $|\alpha|$ in the range of $[0.68, 0.72]$, ultimately selecting $|\alpha| = 0.71$. We want to highlight that in case the real channel loss behaves differently from the assumed loss during the optimization, this leads only to slightly suboptimal $|\alpha|$ values,  leading to lower secure key rates. However, this does not imply any channel loss assumptions for the reported key rates. Thus, the reported key rates are still reliable and independent of any channel model.

The parameters for the Energy Test were carefully selected, considering the complex interplay between different parameters (for more details see Supplementary Note \ref{APDX:secProofDetails}). These choices are detailed in  Table~\ref{tab:generalParams}. The remaining free parameters are the testing ratio $r_T := \frac{k_T}{N}$ and the postselection parameter $\Delta_r$, which we discuss in what follows.

\begin{table}[t]
\centering
\caption{\textbf{Security parameters of involved (sub-) protocols}.}
\resizebox{0.78\hsize}{!}{
\begin{tabular}{|ccc|}
\hline
\multicolumn{1}{|c|}{\textbf{(Sub-) Routine}}                 & \multicolumn{1}{c|}{\textbf{Symbol}}     & \textbf{Value}                 \\ \hline
\multicolumn{1}{|c|}{\textit{\textbf{QKD Protocol}}}          & \multicolumn{1}{c|}{$\epsilon$}             & $\leq 1 \times 10^{-10}$            \\ \hline
\multicolumn{3}{|c|}{\textit{\textbf{}}}                                                                                                     \\ \hline
\multicolumn{1}{|c|}{\textit{\textbf{Privacy Amplification}}} & \multicolumn{1}{c|}{$\epsilon_{\text{PA}}$} & $4 \times 10^{-15}$ \\ \hline
\multicolumn{1}{|c|}{\textit{\textbf{Error Correction}}}    & \multicolumn{1}{c|}{$\epsilon_{\text{EC}}$} & $ \leq\frac{1}{20} \times 10^{-10}$ \\ \hline
\multicolumn{1}{|c|}{\textit{\textbf{Energy Test}}}           & \multicolumn{1}{c|}{$\epsilon_{\text{ET}}$} & $\frac{1}{10} \times 10^{-10}$ \\ \hline
\multicolumn{1}{|c|}{\textit{\textbf{Acceptance Test}}}       & \multicolumn{1}{c|}{$\epsilon_{\text{AT}}$} & $\frac{8}{10} \times 10^{-10}$ \\ \hline
\multicolumn{1}{|c|}{\textit{\textbf{ Smoothing}}}      & \multicolumn{1}{c|}{$\bar{\epsilon}$}     & $\frac{8}{10} \times 10^{-10}$ \\ \hline
\multicolumn{1}{|c|}{\textit{\textbf{ Random Number Generation}}}      & \multicolumn{1}{c|}{$\epsilon_{\text{RNG}}$}     & $\frac{1}{20} \times 10^{-10}$ \\ \hline
\end{tabular}
}
\label{tab:secParams}
\end{table}

\begin{table}[ht]
\centering
\caption{\textbf{Protocol parameter choices  \& experimental parameters}.}
\resizebox{0.98\hsize}{!}{
\begin{tabular}{|c|c|c|}
\hline
\textbf{Parameter}                    & \textbf{Symbol}     & \textbf{Value}                                     \\ \hline
\textit{\textbf{Verification hash length}} & $b_{\mathrm{EV}}$            & $96$ (bit)                                            \\ \hline
$\left\lceil 2\log_2\left( \frac{1}{\epsilon_{\mathrm{PA}}} \right) \right\rceil$ & $b_{\text{PA}}$            & $96$ (bit)                                        \\ \hline
\textit{\textbf{Coherent state amplitude}} & $|\alpha|$            & $0.71$                                            \\ \hline
\textit{\textbf{Cutoff number}}            & $n_c$               & $20$                                               \\ \hline
\textit{\textbf{Detection limit}}          & $M$                 & $5.5$ (NU)                                             \\ \hline
\textit{\textbf{ET - parameter}}           & $\beta_{\mathrm{ET}}$ & $5.5$                                              \\ \hline
\textit{\textbf{Testing ratio}}           & $r_T$ & $\{40\%, 45\%, 50\%\}$                                              \\ \hline
\textit{\textbf{Fraction of outliers}}     & $\frac{l_T}{k_T}$   & $10^{-8}$                                          \\ \hline
\textit{\textbf{Weight}}                   & $w$                 & $\left[1 \times 10^{-7}, 3 \times 10^{-7} \right]$ \\ \hline
\textit{\textbf{$t$-factor}}                   & $t_F$                 & $1$ \\ \hline
\textit{\textbf{Detection efficiency}}                   & $\eta_D$                 & $0.6858$ \\ \hline
\textit{\textbf{Electronic noise}}                   & $\nu_{\mathrm{el}}$                 & $0.0193$ \\ \hline
\textit{\textbf{Est. channel transmittance}}                   & $\eta_{\mathrm{Ch}}$                 & $0.2764$ \\ \hline
\end{tabular}
}
\label{tab:generalParams}
\end{table}

\begin{table*}[ht]
   \centering
\caption{\textbf{Recent advances in DM CVQKD}} \label{tab:table2}
\resizebox{1\hsize}{!}{
\begin{tabular}{|c|c|c|c|c|c|c|c|c|c|}
%         \hline  \multicolumn{3}{|c|}{\textbf{Entry}}  \\
         \hline \bf Ref.  &\bf \makecell{Study \\type}  & \bf \makecell{Constellation \\ order}  & \bf Security & \bf \makecell{Attack\\level}&\bf \makecell{Imperfect\\ device assumption} &\bf Post-processing & \bf Distance, km  & \bf \makecell{Symbol rate,\\ GBaud} &\bf \makecell{SKF, \\bits/symbol$~\times10^{-3}$}\\
     
         \hline  \cite{upadhyaya2021dimension} & Theoretical & General & Asymptotic & Collective &  Trusted noise and loss& - & - & -& -\\
         \hline  \cite{denys2021explicit} & Theoretical &  General& Asymptotic &  Collective  &\makecell{Realistic source}& - & -& -& -\\
         \hline  \cite{matsuura2021finite} & Theoretical & 2 & Composable finite-size & General & Ideal & - & -& -& -\\
         \hline  \cite{lupo2022quantum} & Theoretical & General & Composable finite-size & Collective & \makecell{Realistic detection: finite range,\\ discretization} & - & - & -& -\\
         \hline  \cite{Kanitschar2023} & Theoretical & General & Composable finite-size & Collective& \makecell{Realistic detection: finite range, \\trusted noise and loss} & -& -& -& -\\

        \hline  \cite{roumestan2024shaped} & Experimental & 64, 256 & Asymptotic & Collective & Realistic detection: trusted noise and loss &  Not included & 9.5  & 0.6 & 150.5, 229.5\\
        \hline  \cite{wang2022sub} & Experimental & 4 & Asymptotic & Collective & Realistic detection: trusted noise and loss&  Included & 25 & 5& 10.5\\

        \hline  \cite{tian2023high} & Experimental & 16 & Asymptotic & Collective & Realistic detection:trusted noise and loss &  Not included & 25, 50, 80 & 2.5 &19.6,4.7,0.8\\

        \hline  \cite{pan2022experimental} & Experimental & 64, 256 & Asymptotic & Collective & Realistic detection:trusted noise and loss&  Not included & 50 & 1 &7.6, 9.2\\

        \hline  \cite{hajomer2023continuous} & Experimental & 16, 64 & Asymptotic & Collective & Realistic detection: trusted noise and loss&  Not included & 5 & 8 &21.0, 93.0\\

        \hline This work & Experimental & 4 & Composable finite-size  & Collective & \makecell{Realistic detection: finite range, \\trusted noise and loss}&   Included & 20 & 0.125 &  11.04
\\

         \hline
\end{tabular}
\label{tab:perivouswork}
}
\end{table*}

Next, it is important to clarify how secure key rates were obtained in practice. The length of the raw key that needs to be hashed is given by Eq. (\ref{eq:KRexpression}). We reformulated the right-hand side of Eq. (\ref{eq:KRexpression}) as follows
\begin{equation}\label{eq:KRexpressionPractical}
\begin{aligned}
    \frac{\ell}{N} \leq \frac{n}{N} &\left[ \min_{\rho \in \mathcal{S}^{\mathrm{E\&A}}} H(X|E')_{\rho}- \Delta(w) - \delta(\bar{\epsilon})  - \mathrm{EC}_{\mathrm{leak}}\right]  \\
    & -\frac{1}{N}\left[ n_{\mathrm{blocks}} b_{\mathrm{hash}} + b_{\mathrm{PA}}\right].
\end{aligned}
\end{equation}
The first three terms are obtained from theory, with $\min_{\rho \in \mathcal{S}^{\mathrm{E\&A}}} H(X|E')_{\rho}- \Delta(w)$  obtained from the optimization based on the pre-determined acceptance set, and $\delta(\bar{\epsilon})$ being a correction term related to the entropy chosen in the theoretical analysis (see Theorem \ref{thm:SecurityStatement}). The error-correction leakage $\text{EC}_{\text{leak}}$ normalized per round is given in Eq. (\ref{eq:ECLeak}) and is obtained from the performed error-correction module and discussed in detail in Section \ref{sec:Implementation}. Finally, $n_{\text{blocks}}$ represents the number of blocks into which the uncorrected bit string is divided during error correction, $b_{\text{hash}}$ is the length of the verification hash tag used to verify the correctness of each of the corrected blocks and $b_{\text{PA}}=\left\lceil 2\log_2\left( \frac{1}{\epsilon_{\text{PA}}} \right) \right\rceil$. The chosen values can be found in Table~\ref{tab:generalParams}.

As previously discussed, we began by characterizing the system's honest behavior. This characterization allowed us to define the acceptance set for the honest implementation (see Theorem \ref{Thm:ParameterEstimation}). We selected a $t$-factor of $t_F = 1$, which expands the set of accepted statistics by $\mu$ around each observable. Using this acceptance set, we then calculated the secure key rates for the expected data. For future protocol runs, the procedure simplifies to comparing the observed statistics against this predefined acceptance set. If the observations fall within the set, the entropy term in Eq. (\ref{eq:KRexpression}) is immediately determined, allowing us to proceed with classical postprocessing to obtain secure key rates. Otherwise, the protocol is aborted. Figure \ref{fig:Results_rPS} illustrates the secure key fractions (SKF) obtained upon acceptance as a function of the postselection parameter $\Delta_r$ (ranging from 0 to 0.70), applying the radial postselection strategy~\cite{Kanitschar2021a} for three different testing ratios, $r_T \in \{40\%, 45\%, 50\%\}$.

The experimental SKFs are represented by orange dots, with attached balloons indicating the reconciliation efficiency of the corresponding LDPC codes. The color of the balloons reflects the code rate (R). For comparison, a theoretical curve assuming a $92\%$ error-correction efficiency is plotted (green line). We observe that both the theoretical and experimental SKFs increase with the testing ratio, $r_T$, without yet reaching the point where further increases in $r_T$ reduce the SKF. This effect is observed and discussed in Ref. \cite{Kanitschar2023}. However, due to numerical instabilities for the observed data, we could not go beyond $r_T = 50\%$, reaching the aforementioned inflection point. Postselection is shown to significantly enhance SKF, with $\Delta_r = 0$ representing the absence of postselection. The theoretical curve suggests an optimal SKF at $\Delta_r = 0.60$, assuming a constant reconciliation efficiency. However, the experimental data present a more complex picture. This is because a limited number of LPDC codes, coupled with varying SNR, led to differences in reconciliation efficiency. However, since postselection improves the SNR, potentially allowing for the use of more efficient codes, it exerts a further positive effect on SKF. Additionally, we want to highlight that more postselection eliminates a higher number of signals (see Ref. \cite{Kanitschar2023} for a more detailed discussion of this effect), reducing the demands on classical postprocessing which can be of interest in optimized commercial implementations.

\section{DISCUSSION}
A practical QKD system must meet the requirement of universal composability to ensure that any cryptographic application utilizing the system remains secure. Furthermore, compatibility with standard telecommunication technology is essential for enabling the large-scale deployment of secure quantum key distribution networks. In this study, we provide experimental evidence of a practical DM CVQKD system that successfully distributes composable cryptographic keys, secure against collective attacks, while demonstrating high compatibility with high-speed wireline telecom components.

For error correction, QPSK symbols can be treated as two independent binary symbols. If Gaussian or non-uniform discrete constellations  (e.g., probabilistically shaped 256-quadrature amplitude modulation (256-QAM)) are binary encoded, the bits in their binary encoding are statistically dependent, or in other words correlated. 
Therefore, error correction with high efficiency is easier to perform for QPSK symbols where a single binary LDPC code can be used. Efficient error correcting schemes for correlated bits are more complex to design and to implement, and have higher computational requirements. 

The successful generation of a positive composable key length was achieved through meticulous characterization and optimization of system parameters. This was supported by a high transmission rate of 125 MBaud for coherent states and highly stable system operation—both critical factors in ensuring the system’s overall performance and security. From a theoretical perspective, utilizing the security proof method from Ref.~\cite{Kanitschar2023} offers several advantages for our implementation. First, the resulting lower bounds on the secure key rate are both tight and reliable, accounting for numerical imprecisions, without relying on any assumptions on the channel behavior. Second, the ability to post-select specific symbols introduces additional flexibility, significantly improving key rates. Finally, the numerical approach enables us to incorporate the precisely measured imperfect constellation into the calculations, without relying on unjustified symmetry assumptions, elevating the achieved security claim to a new level. 

Table \ref{tab:perivouswork} highlights the recent theoretical and experimental advancements in DM CVQKD. Notably, discrete modulation allows CVQKD systems to operate at repetition rates comparable to those of classical telecom systems. While security proofs considering composability, general constellations, and realistic assumptions about device imperfections are available, most experimental demonstrations have focused on achieving security against collective attacks in the asymptotic regime, often without fully implementing the post-processing steps. In contrast, this work not only demonstrates composable key distribution but also considers the complete protocol implementation and device imperfections, resulting in key material ready for any cryptographic task. 

Although our work significantly narrows the gap between theoretical and practical implementations, there remains substantial room for further improvement. One key area for development is leveraging the full potential of high-speed wireline components to increase the system's symbol rate to multi-Gbaud. Migrating to a higher modulation format, such as 16-QAM, which is often used in high-speed coherent transceivers, can significantly improve the secure key rate. Additionally, exploiting the advantages of discrete modulation in error-correction could enable high-speed, real-time implementation, which should be the focus of future investigations. As a step toward scalable quantum key distribution networks, multi-user DM CVQKD is another important area to consider, particularly in light of recent theoretical advancements in CV multi-user QKD~\cite{Hajomer2024,bian2023high, kanitschar2024security}.

 To this end, these advancements would further enhance the practicality, cost-effectiveness, and security of real-time, ultra-high-rate QKD systems, paving the way for the large-scale deployment of quantum-safe communication.

\begin{backmatter}
\vspace{0.5cm}
\bmsection{Data availability} All data needed to evaluate the conclusions in this paper are present in the paper and/or the Supplementary Materials. The underlying code for the security argument will be made publicly available within the frame of OpenQKDSecurity \cite{OpenQKDSecurity}.

\smallskip

\bmsection{Acknowledgments} This project was funded within the QuantERA II Programme (project CVSTAR) that has received funding from the European Union’s Horizon 2020 research and innovation program under Grant Agreement No  101017733, Innovation Fund Denmark (IFD)  under Grant Agreement No 731473 and the Austrian Research Promotion Agency (FFG), project number FO999891361; from the European Union’s Digital Europe programme under Grant Agreement No 101091659 (QCI.DK); from the European Union’s Horizon Europe research and innovation programme under the project ``Quantum Security Networks Partnership'' (QSNP, grant agreement no. 101114043). A.A.E.H., U.L.A., and T.G. acknowledge support from Innovation Fund Denmark (CryptQ, 0175-00018A) and the Danish National Research Foundation, Center for Macroscopic Quantum States (bigQ, DNRF142).  A.A.E.H., R.Z., and T.G. acknowledge funding from the Carlsberg Foundation, project CF21-0466. 

\smallskip

\bmsection{Competing interests} The authors declare no competing interests.

\smallskip

\bmsection{Author contributions statement} A.A.E.H. performed the experiment and the overall data processing with assistance from N.J. and R.Z. F.K. conducted the security analysis and secure key calculation under the supervision of C.P. and N.L. F.K. developed and implemented the interface between experiment and theory, including testing routines and data analysis; N.J. provided valuable feedback and suggested improvements. F.K., M.H., and C.P. discussed the application of post-processing to the present data. M.H. implemented and executed the post-processing framework. A.A.E.H., F.K., and M.H. drafted the manuscript with input from all authors. A.A.E.H. and T.G. conceived the experiment. U.L.A., C.P., and T.G. supervised the project. All authors participated in discussions and contributed to the interpretation of the results. 
\smallskip

\end{backmatter}

\bigskip

\bibliography{lib}

\bibliographyfullrefs{lib}

\pagebreak
%\widetext
\onecolumn
\begin{center}
\textbf{\huge Supplementary Materials for\\ \vspace{0.5cm} Experimental composable key distribution using discrete-modulated continuous
variable quantum cryptography}\\ \vspace{0.5cm} Adnan A.E. Hajomer$^{*, \dagger}$, Florian Kanitschar$^{\dagger}$, Nitin Jain, Michael Hentschel, Runjia Zhang, Norbert Lütkenhaus, Ulrik L. Andersen, Christoph Pacher, Tobias Gehring$^{**}$ \\ \vspace{0.5 cm}
$\dagger$ These authors contributed equally\\
Corresponding authors: * aaeha@dtu.dk, ** tobias.gehring@fysik.dtu.dk
\end{center}
\setcounter{equation}{0}
\setcounter{section}{0}
\setcounter{figure}{0}
\setcounter{table}{0}
\setcounter{page}{1}

\makeatletter
\renewcommand{\theequation}{S\arabic{equation}}
\renewcommand{\thefigure}{S\arabic{figure}}
\renewcommand{\thetable}{S\arabic{table}}
\renewcommand{\bibnumfmt}[1]{[S#1]}
\renewcommand{\citenumfont}[1]{S#1}

\section{Application of the security argument}
\phantomsection
\label{APDX:secProofDetails}
In this section, we complement the security proof argument used with additional information. We start by restating the Energy Test theorem.

\begin{theorem}[\textbf{Noise robust Energy Test~\citeS{Kanitschar2023}}\label{thm:energy_test}]
Consider quantum states of the form $\rho^{\otimes N}$, and let $k_T \in \mathbb{N}$, $k_T <\!\!< N$, be the number of signals sacrificed for testing and $l_T \in \mathbb{N}$ be the number of rounds that may not satisfy the testing condition. Denote by $(Y_1, ..., Y_{k_T})$ the absolute values of the results of the test measurement. Pick a weight $w\in [0,1]$, a photon cutoff number $n_{c}$ and a testing parameter $\beta_{\mathrm{test}}$ satisfying $M \geq \beta_{\mathrm{test}} > 0$, where $M>0$ is the finite detection range of the heterodyne detectors.  Define $r:= \frac{\Gamma(n_c+1,0)}{\Gamma(n_c+1,\beta_{\mathrm{test}})}$, where, $\Gamma(n,a)$ is the upper incomplete gamma function, as well as  $Q_y := \begin{pmatrix}
    1-y \\ y
\end{pmatrix}$
 and $P_{j} := \begin{pmatrix}
    1-\frac{j}{k_T}\\
    \frac{j}{k_T}
\end{pmatrix}$. Finally, let $\Pi^{\perp}$ be the projector onto the complement of the photon cutoff space $\mathcal{H}^{n_c}$.\\
Then, as long as $\frac{l_T}{k_T} < \frac{w}{r}$ for all $\rho$ such that $\Tr{ \Pi^{\perp} \rho} \geq w$,
\begin{equation}\label{eq:EnergyTest}
\begin{aligned}
    \mathrm{Pr}&\left[\left|\left\{ Y_j:~ Y_j < \beta_{\mathrm{test}}\right\} \right| \leq l_T \right]  \\ ~~~~& \leq (l_T+1) \cdot 2^{-k_T D\left(P_{l_T} || Q_{\frac{w}{r}}\right)}  =: \epsilon_{\mathrm{ET}},
\end{aligned}
\end{equation}
%\begin{align*}
%    \mathrm{Pr}&\left[\max_{i = 1,...,k_T} Y_i < \beta_{\mathrm{test}}^2  \mathrm{ for all but $l_T$ symbols} \land \mathrm{Tr}[ \Pi^{\perp} \rho] \geq w\right]  \\ & \leq \frac{\left(1-\frac{w}{r} \right)^{k_T-l_T+1}}{k_T-l_T+1} =: \epsilon_{\mathrm{ET}},
%\end{align*}
where $D(\cdot||\cdot)$ is the Kullback-Leibler divergence.
\end{theorem}

Note that due to the nature of the Energy Test, the testing parameters have to be chosen before the protocol execution. The result of the test then is either a pass or fail. In the latter case the protocol aborts. Thus, the choice of the parameters is crucial and non-trivial. Before we come to our particular choices, let us briefly discuss the idea behind the Energy Test to motivate our choices. Quantum states in CVQKD protocols live in infinite-dimensional Hilbert spaces. While this is not a problem per se, many modern security-proof techniques, such as the method used for the present work, employ numerical convex optimization routines. This requires the representation of the objective optimization problem in finite dimensions, hence the clean and rigorous justification of a numerical cutoff, and is the main motivation for conducting the Energy Test. Ref.~\citeS{Kanitschar2023} combines Energy Testing with the Dimension Reduction Method~\citeS{Upadhyaya2021}, which allows relating the infinite-dimensional optimization problem with a finite-dimensional version at the cost of introducing a correction term $\Delta(w)$ that depends on the weight chosen in the Energy Test. While small weights $w$ obviously directly benefit the key rate, the practical choice is a complex interplay between various parameters such as the numerical cutoff $n_c$, which is limited by computational constraints and the testing parameter $\beta_{\text{test}}$ as well as the fraction of allowed outliers, which depend on the physical quantum channel. Finally, all together influence the security parameter $\epsilon_{\text{ET}}$ and the abort probability of the Energy Test. Thus, the overarching goal is to find an as small as possible $w$ together with a computationally feasible $n_c$ that leads to the required security parameter $\epsilon_{\text{ET}}$ with low abort probability for the Energy Test.

Usually, implementations of QKD systems aim for certain total security parameters $\epsilon$, which in turn is a function of the security epsilons of all involved sub-protocols. Therefore, based on considerations about scaling and the impact of the security parameters on the key rate, usually, the security parameter $\epsilon_{\text{ET}}$ is fixed, too. The numerical cutoff $n_c$ may be limited by available computational capabilities. Within these limitations, usually, the cutoff number is chosen as a compromise between calculation time and the expected success rate of the Energy Test. Based on experience from theory work \cite{Kanitschar2023}, we chose $n_c = 20$ for this work. To avoid inefficiencies, we chose the detection limit $M$ equal to the testing parameter $\beta_{\text{test}}$. The numerical value of $M = 5.25$ was chosen based on expected honest channel behavior for the given system parameters and the coherent state amplitude used. We fix the number of allowed outliers $l_T$ as a fraction of the total number of test symbols $k_T$ again based on the expected behavior of the system. Since, according to Eq. (\ref{eq:EnergyTest}), the $\epsilon_{\text{ET}}, n_c, \beta_{\text{test}}, l_T$ and the weight $w$ are related via a non-linear function, we then search for the smallest $w$ for fixed parameters $n_c, \beta_{\text{test}}$ and $ l_T$ such that $\mathrm{Pr}\left[\left|\left\{ Y_j:~ Y_j < \beta_{\mathrm{test}}\right\} \right| \leq l_T \right]$ is smaller or equal to the pre-defined $\epsilon_{\text{ET}}$, using a bisection-method. Thus, the actual security parameter for the Energy Test usually is slightly smaller than the required $\epsilon_{\text{ET}}$. Hence, we fixed all testing parameters while still guaranteeing that the subroutine obeys the security claim.

Next, we restate the Acceptance Testing theorem, which is based on Hoeffding's inequality.
\begin{theorem}[\textbf{Acceptance Test }\cite{Kanitschar2023}]\label{Thm:ParameterEstimation}

Let $\Theta$ be the set of Bob's observables. Let $\mathbf{r} \in \mathbb{R}^{|\Theta|}$ and $\mathbf{t} \in \mathbb{R}^{|\Theta|}_{\geq 0}$, where $|\Theta|$ denotes the cardinality of $\Theta$. Define the set of accepted statistics as
\begin{equation}\label{eq:accepted-observations} \mathcal{O} := \{ \mathbf{v} \in \mathbb{R}^{\Theta} : \forall X \in \Theta, |v_{X} - r_{X}| \leq t_{X} \} \ ,
\end{equation}
and the corresponding acceptance set as
\begin{equation}\label{eq:ATset}
\begin{aligned}
    &\mathcal{S}^{\mathrm{AT}}:=  \left\{ \rho \in \mathcal{D}(\mathcal{H}_A \otimes \mathcal{H}_B^{n_c}):\right. \\
    & \hspace{18mm} \forall X \in \Theta, |\Tr{\rho X} - r_{X}| \leq \mu_{X} + t_{X} \},
\end{aligned}
\end{equation}
where $r_{X}$ is the $X$th element of the vector $\mathbf{r}$ and likewise for $t_{X}$. For every $X \in \Theta$, let \begin{equation*}
    \mu_X := \sqrt{\frac{2x^2}{m_{X}} \ln\left( \frac{2}{\epsilon_{\mathrm{AT}}} \right)}
\end{equation*}
or, if $X$ is a positive semidefinite operator,
\begin{equation*}
    \mu_X := \sqrt{\frac{x^2}{2m_{X}} \ln\left( \frac{2}{\epsilon_{\mathrm{AT}}} \right)},
\end{equation*}
where $x := \|X\|_{\infty}$ and $m_{X}$ is the number of tests for the observable $X$. If $\rho \not \in \mathcal{S}^{\mathrm{AT}}$, then the probability of accepting the statistics generated by the i.i.d. measurements of $\rho^{\otimes n}$ is bounded above by $\epsilon_{\mathrm{AT}}$. That is, the complement of $\mathcal{S}^{\mathrm{AT}}$ are all $\epsilon_{\mathrm{AT}}$-filtered.
\end{theorem}

Before we can give further details, we need to specify the observable used. In accordance with \citeS{Upadhyaya2021, Kanitschar2023}, we chose the displaced photon number operator $\hat{n}_{\beta_i}$ and the displaced squared photon number operator $\hat{n}_{\beta_i}^2$, where $\beta_i \in \{\sqrt{\eta} e^{\frac{i \pi}{4}} |\alpha|, \sqrt{\eta} |\alpha| e^{\frac{3i \pi}{4}}, \sqrt{\eta} |\alpha| e^{\frac{5i \pi}{4}}, i\sqrt{\eta} |\alpha| e^{\frac{7i \pi}{4}}\}$, for $\eta$ an experimentally determined estimator of the total loss in the system. Thanks to the finite detection range, these operators can be bounded as follows $||\hat{n}_{\beta_j}||_{\infty} = M^2-\frac{1}{2}$ and $||\hat{n}_{\beta_j}^2||_{\infty} = M^4-\frac{1}{2}M^2$, which leads to $\mu_{\hat{n}_{\beta_j}} = \sqrt{\frac{||\hat{n}_{\beta_j}||_{\infty}^2}{2k_T} \ln\left( \frac{2}{\epsilon_{\textrm{AT}}} \right)}$ and $\mu_{\hat{n}_{\beta_j}^2} = \sqrt{\frac{||\hat{n}_{\beta_j}^2||_{\infty}^2}{2k_T} \ln\left( \frac{2}{\epsilon_{\textrm{AT}}} \right)}$. Note that lowering the security parameter of the Acceptance Test increases the corresponding $\mu$ for a constant number of test symbols $m$. Hence, small $\epsilon_{\text{AT}}$ usually comes with a considerable cost for the secure key rate. Therefore, when adjusting the epsilons among the involved sub-routines, one usually tries to choose $\epsilon_{\text{AT}}$ as large as possible. Additionally, note that, in accordance to Ref. \citeS{Kanitschar2023}, we introduce $t_F$ to describe the parameter $t_X$ in the non-unique acceptance scenario as factor measuring $t_X$in multiples of $\mu_X$, $t_X = t_F \mu_X$.

Thus, the optimization problem $\min_{\rho \in \mathcal{S}^{\text{E\&A}}} f(\rho)$ for $f(\rho) := H(X|E')_{\rho}$ is given by the following semi-definite program \cite{Kanitschar2023}
\begin{align}
\begin{aligned}\label{eq:PrimalProblem}
    \beta :=& \min~ f(\bar{\rho})\\
    \text{s.t. }& \\
    & \Tr{P}+\Tr{N} \leq 2 \sqrt{w}\\
    & P \geq \mathrm{Tr}_{B}\left[ \overline{\rho} \right] - \rho_A\\
    & N \geq - \left( \mathrm{Tr}_{B}\left[ \bar{\rho} \right] - \rho_A \right) \\
    &\Tr{\left( \ket{j}\!\bra{j}\otimes\hat{n}_{\beta_j}\right) \bar{\rho} } \geq \mu_{\hat{n}_{\beta_j}} + \langle \hat{n}_{\beta_j} \rangle - w ||\hat{n}_{\beta_j}||_{\infty} \\
    & \Tr{\left( \ket{j}\!\bra{j}\otimes\hat{n}_{\beta_j}\right) \bar{\rho} } \leq \mu_{\hat{n}_{\beta_j}} + \langle \hat{n}_{\beta_j} \rangle  \\
    &\Tr{\left( \ket{j}\!\bra{j}\otimes\hat{n}_{\beta_j}\right) \bar{\rho}} \geq -\mu_{\hat{n}_{\beta_j}} + \langle \hat{n}_{\beta_j} \rangle - w ||\hat{n}_{\beta_j}||_{\infty} \\
    & \Tr{\left( \ket{j}\!\bra{j}\otimes\hat{n}_{\beta_j}\right) \bar{\rho}} \leq -\mu_{\hat{n}_{\beta_j}} + \langle \hat{n}_{\beta_j} \rangle  \\ 
    & \Tr{\left( \ket{j}\!\bra{j}\otimes \hat{n}^2_{\beta_j} \right) \bar{\rho}}\geq \mu_{\hat{n}^2_{\beta_j}} + \langle \hat{n}_{\beta_j}^2 \rangle - w ||\hat{n}_{\beta_j}^2||_{\infty}\\
    &\Tr{\left( \ket{j}\!\bra{j}\otimes \hat{n}^2_{\beta_j} \right) \bar{\rho}} \leq \mu_{\hat{n}^2_{\beta_j}} + \langle \hat{n}^2_{\beta_j} \rangle \\
    & \Tr{\left( \ket{j}\!\bra{j}\otimes\hat{n}^2_{\beta_j}\right) \bar{\rho} } \geq -\mu_{\hat{n}^2_{\beta_j}} + \langle \hat{n}_{\beta_j}^2 \rangle - w ||\hat{n}_{\beta_j}^2||_{\infty} \\
    & \Tr{\left( \ket{j}\!\bra{j}\otimes\hat{n}^2_{\beta_j}\right) \bar{\rho} } \leq -\mu_{\hat{n}^2_{\beta_j}} + \langle \hat{n}^2_{\beta_j} \rangle  \\ 
    & 1-w \leq \Tr{\overline{\rho}} \leq 1\\
    & \bar{\rho}, P, N \geq 0
\end{aligned}
\end{align}
where $j \in \{0,..., 3\}$. As already mentioned in the main text, this problem can be solved with the numerical method developed in Refs. \citeS{Coles2016, Winick2018} using the Frank-Wolfe algorithm \citeS{FrankWolfe1956}.

\section{Data Analysis}
\phantomsection
\label{APDX:DataAnalysis} 
\subsection{Setting the Stage}
In this section, we aim to bridge the gap between experimental observations and theoretical key rate calculations, discussing the data analysis conducted for the Energy Test and Acceptance Test (Steps 3 and 4 in the protocol detailed in \citeS{Kanitschar2023}). Additional theoretical background is available in Ref.~\citeS{UpadhyayaThesis2021}, where the analysis for the asymptotic regime is discussed. The coding was performed using \textsc{Matlab}\textsuperscript{\textregistered}, version 2022a.

Each coherent detection comprising the heterodyne detection in Bob's lab outputs a number corresponding to the result of the analog-to-digital converter (ADC). We denote the pair of numbers representing the output of the heterodyne detection by $(\tilde{y}^q_k,\tilde{y}^p_k)$ for each round $k$. However, these numbers currently lack a specific meaning and depend on the characteristics of the detector, which are unknown. Thus, the first step is to measure the variance of the shot noise, yielding another pair of numbers $V_{\text{shot}}(q)$ and $V_{\text{shot}}(p)$. These numbers can then be used to interpret our measurement results in multiples of the square root of the shot noise: $q_{\text{SNU}} = \frac{\tilde{y}k^q}{\sqrt{V{\text{shot}}(q)}}$ and $p_{\text{SNU}} = \frac{\tilde{y}k^p}{\sqrt{V{\text{shot}}(p)}}$. This is known as shot-noise units (SNU). While this unit system, by definition, satisfies the uncertainty relation $\sigma_q \sigma_p \geq 1$, the relationship between its corresponding ladder operators and quadrature operators is asymmetric. This can be corrected by switching to natural units: $q_{\text{NU}} = \frac{q_{\text{SNU}}}{\sqrt{2}}$ and $p_{\text{NU}} = \frac{p_{\text{SNU}}}{\sqrt{2}}$, which is the unit system used in our analysis. The rescaled uncertainty relation is $\sigma_q \sigma_p \geq \frac{1}{2}$, and the corresponding quadrature operators are related to the ladder operators via $\hat{q}_{\text{NU}} = \frac{\hat{a}^{\dagger} + \hat{a}}{\sqrt{2}}$ and $\hat{p}_{\text{NU}} = i \frac{\hat{a}^{\dagger} - \hat{a}}{\sqrt{2}}$. Consequently, the photon number operator is $\hat{n} = \frac{1}{2}\left( \hat{q}_{\text{NU}}^2 + \hat{p}_{\text{NU}}^2 - 1 \right)$. In natural units, the coherent state parameter $\alpha = \alpha_r + i \alpha_i$ can be related to the expectations of quadrature measurements:
\begin{align}
\begin{aligned}\label{eq:APDX:relationQP_Alpha}
    \hat{a} \ket{\alpha} &= \alpha \ket{\alpha} \\
    \frac{\hat{q}_{\text{NU}} + i \hat{p}_{\text{NU}}}{\sqrt{2}} \ket{\alpha} & = \alpha \ket{\alpha} \\
    \Rightarrow \frac{\langle \hat{q}_{\text{NU}}\rangle_{\ket{\alpha}} + i \langle \hat{p}_{\text{NU}}\rangle_{\ket{\alpha}}}{\sqrt{2}} &= \alpha = \alpha_r + i \alpha_i,
\end{aligned}
\end{align}
where $\ket{\alpha}$ is an eigenstate of the annihilation operator. This information is crucial when applying postselection and the Energy Test.

To ease notation, from now on, we assume that $q$ and $p$ are given in natural units and omit the subscript $\text{NU}$, $q \equiv q_{\text{NU}}$ and  $p \equiv p_{\text{NU}}$.

\subsection{Analysis of the Experimental Data}
In our experiment, the chosen coherent state amplitude, as well as the probability distribution (uniform distribution) for the quantum states are known. Additionally, the detector parameters $\eta_D$ and $\nu_{\text{el}}$ can be determined by back-to-back measurements and are considered known within experimental accuracy. The remaining task is to analyze the data produced by the heterodyne detection.

A random subset of size $k_T$ of the recorded data is read out, normalized to shot noise, and transformed into natural units. While the heterodyne detection determines values for the $q$ and $p$ quadrature for each round, as outlined in Section \ref{sec:Theory}, the considered observables are displaced versions of the photon number, $\hat{n}_{\beta_i}$, and the squared photon number, $\hat{n}_{\beta_i}^2$, with displacement $\beta_i = \sqrt{\eta} \alpha_i$. Intuitively, if there was no noise, the total transmittance $\eta$ was known perfectly and the channel was perfectly Gaussian, the expectation for those observables would be $0$. Thus, the deviation from $0$ `quantifies' the aberration from this idealized channel in some sense.

Therefore, our first step is calculating the average $q$ and $p$ value for each of the four constellation points separately, followed by displacing each of the measured symbols accordingly. Let $q_k^{(i)}$ and $p_k^{(i)}$ denote the measurement results in each round for constellation point $i$ and let $\bar{q}^{(i)} = \frac{1}{k_T^{(i)}} \sum_{j=1}^{k_T^{(i)}} q_i$ and $\bar{p}^{(i)} = \frac{1}{k_T^{(i)}} \sum_{j=1}^{k_T^{(i)}} p_i$ be the averages for each constellation point, where $k_T^{(i)}$ is the number of symbols where $\alpha_i$ was prepared. We obtain displaced quantities $\Tilde{q}^{(i)}_k:=q_k^{(i)} - \bar{q}^{(i)}$ and $\Tilde{p}^{(i)}_k:=p_k^{(i)} - \bar{p}^{(i)}$ and calculate the noisy versions of the expectations of $\hat{n}_{\beta_i}$ and $\hat{n}_{\beta_i}^2$ via (see Ref.~\citeS{UpadhyayaThesis2021} or derive directly)
\begin{align}
    \bar{n}_i^{\text{nsy}} := \frac{1}{k_T^{(i)}} \sum_{k=1}^{k_T^{(i)}} \left[ \frac{1}{2} \left(\tilde{q}_k^{(i)} \right)^2 + \left(\tilde{p}_k^{(i)} \right)^2 -1 \right]
\end{align}
and
\begin{align}
    \bar{n}_i^{\text{sq, nsy}} := \frac{1}{k_T^{(i)}} \sum_{k=1}^{k_T^{(i)}} \left[ \frac{1}{4} \left(\tilde{q}_k^{(i)} \right)^4 + \frac{1}{2} \left(\tilde{q}_k^{(i)} \right)^2  \left(\tilde{p}_k^{(i)} \right)^2 + \frac{1}{4} \left(\tilde{p}_k^{(i)} \right)^4 - \frac{3}{2} \left(\tilde{q}_k^{(i)} \right)^2 - \frac{3}{2} \left(\tilde{p}_k^{(i)} \right)^2 + 1 \right].
\end{align}

In case one wants to make use of the trusted detection noise, the relation between the noisy and trusted observables is given by~\cite{UpadhyayaThesis2021}
\begin{align}
    \bar{n}_i^{\text{tr}} = \frac{\bar{n}_i^{\text{nsy}}-\nu_{\text{el}}}{\eta_D}
\end{align}
and
\begin{align}
    \bar{n}_i^{\text{sq, tr}} = \frac{\bar{n}_i^{\text{sq, nsy}} - 2 \nu_{\text{el}}^2 - \nu_{\text{el}} -\left(4\nu_{\text{el}}+1-\eta_D\right)\left( \bar{n}_i^{\text{nsy}} - \nu_{\text{el}}\right)}{\eta_D^2}.
\end{align}

Those values can now be inserted into the right-hand sides of the constraints in the optimization problem given in Eq. (\ref{eq:PrimalProblem}).\\
It remains to discuss the Energy Test and postselecion. Quantities like the postselection parameter $\Delta_r$ and the detection limit $M$ are given in natural units, making the practical comparison for each key round $k$ straight-forward: 
\begin{align}
    \Delta_r \leq \abs{q_k+ip_k} \leq M.
\end{align}

Contrarily, the testing parameter for the Energy Test $\beta_{\text{test}}$ (see Theorem 2 in \citeS{Kanitschar2023}) is compared to coherent state amplitudes, and thus is dimensionless. Therefore, for the Energy Test, we make use of our findings in Eq. (\ref{eq:APDX:relationQP_Alpha}) and count the number of test symbols $j$ satisfying
\begin{equation}
    \abs{\frac{q_j + i p_j}{\sqrt{2}}} > \beta_{\text{test}}.
\end{equation}
Thus, according to the Energy Testing theorem, if $\left|\left\{ j \in\{1,...,k_T\}:~ \abs{\frac{q_j + i p_j}{\sqrt{2}}} > \beta_{\text{test}} \right\}\right| \leq \ell_T$, this is if the obtained number of outliers is less than or equal to $\ell_T$, the test passes; otherwise, it fails.

Note that in order to choose the combination between testing parameter tight and detection limit tight, we require $\beta_{\text{test}} = \frac{M}{\sqrt{2}}$.

We conclude with a note about the channel loss $\eta_{\text{Ch}}$ and the excess noise $\xi$, which are often reported in the literature and important for practical comparison of different works but play only a secondary role in the security argument used in this work. We use a back-to-back measurement to quantify $\alpha_{\text{B2B}},$ the coherent state amplitude in Alice's lab. Based on Bob's measurements in the test symbols, we determine $\beta_{\text{rec}} := \frac{1}{k_T} \sum_{j=1}^{k_T} \abs{\frac{q_j + i p_j}{\sqrt{2}}}$, which we backpropagate by the earlier quantified detection efficiency, $\beta_{\text{back}} := \frac{\beta_{\text{rec}}}{\sqrt{\eta_D}}$. Then, we obtain an estimate for the channel loss by relating the mean photon numbers of the sent and the received state, $\eta_{\text{Ch}} = \frac{\beta_{\text{back}}^2}{\alpha_{\text{B2B}}^2}$.

This estimated channel loss can be used to determine an estimate for the excess noise $\xi$. Therefore, we take the mean over all four symbols of $\bar{n}_i^{\text{tr}}$, $\bar{n}:= \frac{1}{4} \sum_{i=1}^{4} \bar{n}_i^{\text{tr}}$ and calculate $\xi = \frac{2 \bar{n}}{\eta_{\text{Ch}}}$. The factor of $2$ is to compensate the $50:50$ beamsplitter in the heterodyne detector. 

Alternatively, the channel loss can also be determined separately for each of the four symbols, obtaining $\eta_{\text{Ch}}^{i}$. Then, instead of averaging over $\bar{n}_i^{\text{tr}}$, we can directly calculate $\xi_i = \frac{2 \bar{n}_i^{\text{tr}}}{\eta_{\text{Ch}}^{i}}$. 

However, we note that those values are calculated mainly for comparison reasons with other works, as the security argument works directly with the measurement outcomes $q_j$ and $p_j$ and derived quantities. Therefore, we never actively use any assumption about the channel such as Gaussianity.

 \newpage

\end{document}